\documentclass[aip]{revtex4-1}

\usepackage{graphicx}%
\usepackage{dcolumn}%
\usepackage{bm}%

\newcommand {\be}{\begin{equation}} % start equation
\newcommand{\ee}{\end{equation}}    % end equation

\usepackage{cases}

\begin{document}
\baselineskip 20 pt
\today
\begin{center}
{\LARGE  Gyro-viscosity and linear dispersion relations in  pair-ion magnetized plasmas}
\end{center}

\begin{center}
M. Kono and   J. Vranjes${}^{1, 2}$ \\
Faculty of Policy Studies, Chuo University, Tokyo, Japan\\
${}^{1}$Instituto de Astrofisica de Canarias,  Tenerife, Spain\\
${}^{2}$Departamento de Astrofisica, Universidad de La Laguna,  Tenerife, Spain.
\end{center}

\begin{center}
Abstract
\end{center}

\bigskip

A fluid theory has been developed by taking account of gyro-viscosity to study wave propagation characteristics in a homogeneous pair-ion magnetized plasma with a cylindrical symmetry. The exact dispersion relations derived by the Hankel-Fourier transformation are shown  comparable  with those observed in the experiment by Oohara and co-workers. The gyro-viscosity is responsible for the change in propagation characteristics of the ion cyclotron wave from forward to backward by suppressing the effect of the thermal pressure which normally causes the forward nature of dispersion.
Although the experiment has been already explained by a kinetic theory by the present authors,   the kinetic derivations are so involved because of exact particle orbits in phase space, finite Larmor radius effects, and higher order ion cyclotron resonances. The present fluid theory provides  a simple and transparent structure to the dispersion relations since the gyro-viscosity is renormalized into
the ion cyclotron frequency which itself indicates the backward nature of dispersion. The usual disadvantage of a fluid theory, which treats only fundamental modes of eigen-waves excited in a system and is not able to describe higher harmonics that a kinetic theory does, is compensated by simple derivations and clear picture based on the renormalization of the gyro-viscosity.
\\



\pagebreak

\vspace{0.7cm}

\noindent{\bf I. \,\,\, Introduction}

\vspace{0.7cm}

The pioneering experimental studies on propagation properties of electrostatic waves in pair-ion plasmas$^{1-5}$ % \cite{oh1}-\cite{oh5},
have been explained by a kinetic theory$^{6,7}$, which shows the importance of the finite Larmor radius effects and suggests that a fluid theory can reproduce the experimental results so long as the fundamental waves are concerned if the finite
Larmor radius effects are properly taken into account. The main point of the experiment is backward waves which had been left unexplained after a number of theoretical works$^{8-14}$. The frequency ranges of the main backward waves are discretized as $\omega \le \Omega$ and $3 \Omega \le \omega \le 5 \Omega$, where $\Omega$ is the ion cyclotron frequency. This indicates that the backward waves are related to the higher harmonics of the ion cyclotron waves inherent to a kinetic theory.

Thus the kinetic theory developed by the present authors is  indispensable to understanding the peculiar behavior of the observed dispersion relations.  However the derivations in the kinetic theory are not straightforward, and furthermore the obtained dispersion equation is also complicated because of involvement from  higher order cyclotron resonances. Therefore it is natural to develop a simple fluid theory which grasps the physical essence inherent to the backward nature of the dispersion at the expense of whole pictures given by the kinetic theory.

Certainly a fluid theory cannot describe the higher harmonics of the ion cyclotron wave and is not responsible for the intermediate frequency waves whose frequencies are $ \Omega \le \omega \le 5 \Omega$. However the fundamental ion cyclotron wave is treated by a fluid theory and if it is converted from a forward wave to a backward wave by the finite Larmor radius effects, the intermediate frequency waves as a result of a coupling of higher harmonics of the ion cyclotron wave is expected to be backward even though a fluid theory cannot treat the intermediate frequency waves directly. In fact the kinetic theory$^{6,7}$ shows the intermediate frequency waves are excited by coupling of the even $2m$ and odd $2m+1$ harmonics of the ion cyclotron wave with $m \ge 1$.  Not all the higher harmonics of the ion cyclotron wave are backward.

In the pair-ion plasmas used in the experiments the Larmor radii of positive and negative ions are not different and
their dynamics are the same except for the rotation direction of the cyclotron motion. However the finite Larmor radius brings the difference in the velocity of the neighboring parts of the fluid and the velocity shear drives the momentum transfer through the gyro-viscosity. The gyro-viscosity is  invariant with respect to time reversal in contrast to the collisional viscosity which is not invariant under time reversal. Therefore the gyro-viscosity is included in the real part of the dispersion relation while the collisional viscosity is included in the imaginary part of the dispersion relation. This indicates that the gyro-viscosity may reflect the finite Lamor radius effects to modify the usual mode behavior, and some of the puzzling features observed in the experiments$^{1-5}$ could be explained even within the fluid theory.

In unbounded plasmas the gyro-viscosity coefficients for directions perpendicular and parallel to a magnetic field are given by Braginskii, and they read  $\eta_{\alpha\perp}=v_{T\alpha}^2/2\Omega_{\alpha}$ and $\eta_{\alpha\parallel}=v_{T\alpha}^2/\Omega_{\alpha}$, respectively. Here, $\Omega_{\alpha}=e_{\alpha}B/m_{\alpha}c$ and $v_{T\alpha}^2=T_{\alpha}/m_{\alpha}$ and $\alpha$ denotes the species of the particle. These gyro-viscosity coefficients are obtained under the condition that the Larmor radius is much smaller than the plasma scale length. In the plasma used in the experiments, however, the Larmor radius is around  1/2 of the plasma radius and the conventional gyro-viscosity may be modified. But  we shall assume that the gyro-viscosity coefficients $\eta_{\alpha\perp}$ and $\eta_{\alpha\parallel}$ are applicable to the present problem, and we shall check if the theory meets the experimental results.  In this case the gyro-viscosity is renormalized into the cyclotron frequency
in a way $\Omega_{\alpha}[1-(k_{\perp}^2/2 + k_z^2)\rho_{\alpha}^2]$, where $\rho_{\alpha}=v_{T\alpha}/\Omega_{\alpha}$ is the Larmor radius, and it may suppress  the thermal dispersion term.

It will be shown that the gyro-viscosity converts the ion cyclotron wave from forward to backward.
In a cylindrical coordinate system the Hankel transformation is a key tool to renormalize the gyro-viscosity into the cyclotron frequency and yield an exact
dispersion equation which is readily solved. This is a definite advantage of our fluid theory over the kinetic theory.

In Sec.~II,  starting from a two-fluid model of  singly charged positive and negative ions with the same mass, the dispersion equation is obtained by introducing the Fourier-Hankel transformation. In Sec.~III  the dispersion equation is solved to give dispersion relations which are shown in Sec.~IV to be comparable with the experimental ones of the low and high frequency waves.
Discussion is given in the last section.

\vspace{0.7cm}

\noindent{\bf II. \,\,\, Basic Equations in a Cylindrically Symmetric Homogeneous System}

\vspace{0.7cm}

We start with the basic equations for a two-fluid plasma of singly charged positive and negative  ions immersed in  an external magnetic field in the axial direction of a cylinder. Since the masses of both ions are large, in the following the gyro-viscosity is taken into account but not the collisional viscosity because we may assume that the cyclotron frequencies are large compared with the collision frequency. Our starting fluid equations are:
\begin{eqnarray}
 \frac{\partial n_{\alpha}}{\partial t} +
{\bm \nabla} \cdot (n_{\alpha} {\bm v}_{\alpha}) &=& 0,\label{c201}\\
 \frac{\partial {\bm v}_{\alpha}}{\partial t} +
 ({\bm v}_{\alpha}\cdot {\bm \nabla}) {\bm v}_{\alpha} &=&
\frac{e_{\alpha}}{m_{\alpha}}\left({\bm E}
+\frac{1}{c}{\bm v}_{\alpha}\times {\bm B}\right)
- \frac{1}{m_{\alpha}n_{\alpha}}\left({\bm \nabla} p_{\alpha} + {\bm \nabla}\cdot{\bm \pi_{\alpha}}\right).\label{c202}
\end{eqnarray}
Here, $\alpha$ denotes  positive and negative ions and ${\bm \pi}$ is the gyro-viscosity tensor given by
\begin{eqnarray}
\frac{1}{n_{\alpha}m_{\alpha}}({\bm \nabla}\cdot{\bm \pi})_{\perp} &=& -\left(\eta_{\alpha\perp}{\bm \nabla}_{\perp}^2 + \eta_{\alpha\parallel}\frac{\partial^2}{\partial z^2}\right){\bm v}_{\alpha} \times {\bm z}
- \eta_{\alpha\parallel}({\bm \nabla} \times {\bm z}  )\frac{\partial}{\partial z}v_{\alpha z},\label{c203}\\
\frac{1}{n_{\alpha}m_{\alpha}}({\bm \nabla}\cdot{\bm \pi})_z &=& -\eta_{\alpha\parallel}
\frac{\partial}{\partial z}{\bf z} \cdot({\bm \nabla}\times {\bf v}_{\alpha}),\label{c204}
\end{eqnarray}
where ${\bm z}$ is the unit vector parallel to the magnetic field.
The Poisson equation together with the continuity equation reads as
\begin{equation}
{\bm \nabla}\cdot{\bm D}={\bm \nabla} \cdot \left( \frac{\partial {\bm E}}{\partial t} + 4\pi \sum_{\alpha}e_{\alpha}n_{\alpha}{\bm v}_{\alpha} \right) =0.\label{c205}
\end{equation}
In a cylindrical  system the equation of motion  is rewritten through its three components as
\begin{eqnarray}
&&\frac{\partial v_{\alpha r}}{\partial t} + v_{\alpha r}\frac{\partial v_{\alpha r}}{\partial r} + \frac{v_{\alpha\theta}}{r}\frac{\partial v_{\alpha r}}{\partial \theta} + v_{\alpha z}\frac{\partial v_{\alpha r}}{\partial z} -\frac{v_{\alpha \theta}^2}{r}\nonumber\\
&&= - v_{T\alpha}^2 \frac{\partial}{\partial r}(\varphi_{\alpha}+\ln n_{\alpha})
+ \Omega_{\alpha}\left(1+\eta_{\alpha\perp}\nabla_{\perp}^2+\eta_{\alpha\parallel}\frac{\partial^2}{\partial z^2}\right)v_{\alpha \theta}
+ \eta_{\alpha\parallel}\frac{1}{r}\frac{\partial}{\partial \theta}\frac{\partial v_{\alpha z}}{\partial z},\label{c206}\\
&&\frac{\partial v_{\alpha \theta}}{\partial t} + v_{\alpha r}\frac{\partial v_{\alpha\theta}}{\partial r}+\frac{v_{\alpha\theta}}{r}\frac{\partial v_{\alpha\theta}}{\partial \theta}+v_{\alpha z}\frac{\partial v_{\alpha \theta}}{\partial z} + \frac{v_{\alpha\theta}v_{\alpha r}}{r}\nonumber\\
&&= - v_{T\alpha}^2 \frac{1}{r}\frac{\partial}{\partial \theta}(\varphi_{\alpha}+\ln n_{\alpha})
- \Omega_{\alpha}\left(1+\eta_{\alpha\perp}{\nabla}_{\perp}^2+\eta_{\alpha\parallel}\frac{\partial^2}{\partial z^2}\right)v_{\alpha r}
-  \eta_{\alpha\parallel}\frac{\partial}{\partial r}\frac{\partial v_{\alpha z}}{\partial z},\label{c207}\\
&&\frac{\partial v_{\alpha z}}{\partial t} + v_{\alpha r}\frac{\partial v_{\alpha z}}{\partial r} + \frac{v_{\alpha\theta}}{r}\frac{\partial v_{\alpha z}}{\partial \theta} + v_{\alpha z}\frac{\partial v_{\alpha z}}{\partial z}\nonumber\\
&&= - v_{T\alpha}^2 \frac{\partial}{\partial z}(\varphi_{\alpha}+\ln n_{\alpha})
 + \eta_{\alpha\parallel}\frac{\partial}{\partial z}\left[\frac{1}{r}\frac{\partial}{\partial r}(rv_{\alpha \theta})-\frac{1}{r}\frac{\partial v_{\alpha r}}{\partial \theta}\right],\label{c208}
\end{eqnarray}
where the potential $\phi$ is used instead of the electric field ${\bm E}$ with a normalization
$\varphi_{\alpha}=e_{\alpha}\phi/T_{\alpha}$.
The continuity equation reads
\begin{equation}
\frac{\partial n_{\alpha}}{\partial t} + \frac{1}{r}\frac{\partial}{\partial r}(rn_{\alpha}v_{\alpha r})+\frac{1}{r}\frac{\partial}{\partial \theta}(n_{\alpha}v_{\alpha \theta})+\frac{\partial}{\partial z}
(n_{\alpha}v_{\alpha z})=0.\label{c209}
\end{equation}
It is worthwhile to note that linearized equations eqs.~(6) and (7) are combined to give
\[
i\frac{\partial}{\partial t}(v_{\alpha r}+ iv_{\alpha\theta})=\Omega_{\alpha}(v_{\alpha r}+ iv_{\alpha\theta}) + \Omega_{\alpha}\left(\eta_{\alpha\perp} \nabla_{\perp}^2 +\eta_{\alpha\parallel}\nabla_{\parallel}^2\right)\left(v_{\alpha r}+ iv_{\alpha\theta}\right)
\]
\begin{equation}
+ i v_{T\alpha}^2\left(\frac{\partial}{\partial r} +\frac{1}{r}\frac{\partial}{\partial \theta}\right)\left(\varphi_{\alpha}+\ln n_{\alpha}\right),\label{c210}
\end{equation}
which is a Schrodinger type of wave equation and not a diffusion equation. Thus the gyro-viscosity is irrespective of the dissipative effect.

 \vspace{0.7cm}

\noindent{\bf III. \,\,\, Linear Dispersion Relations}

\vspace{0.7cm}

We consider a  plasma confined in a cylindrical vessel whose radius $R$ is 4 cm and the axial length is 90 cm.
Although the plasma is confined in a narrow vessel we assume that the background density is constant. The external magnetic field is assumed constant as well. In the experiments only longitudinal waves are excited  whose dispersion relations are measured. However we study general cases where  waves propagate in the $z$ direction parallel to the magnetic field as well as in the azimuthal direction and the radial structures have to be determined by eigenvalue equations.

 First we introduce the Hankel transform which is the best tool in a cylindrical coordinate system and converted from the Fourier transform in a Cartesian coordinate system.
\begin{eqnarray}
A({\bm x}) &=& A(r, \theta, z)=\sum_{{\bm k}} A({\bm k})e^{i{\bm k}\cdot{\bm x}}
   = \sum_{{\bm k}_{\perp}, k_z} A({\bm k}_{\perp}, k_z) e^{i({\bm k}_{\perp}\cdot{\bm r}+k_z z )}\nonumber\\
   &=& \sum_{k_{\perp}, k_z}\int \frac{d\rho}{2\pi} A(k_{\perp}, \rho, k_z)e^{i(k_{\perp}r\cos (\theta - \rho) + k_z z)}\nonumber\\
   &=& \sum_{k_{\perp}, k_z}  \int \frac{d\rho}{2\pi}  \sum_{\ell'} A_{\ell'}(k_{\perp}, k_z)e^{i\ell' \rho } \sum_{\ell}J_{\ell}(k_{\perp}r) e^{i\ell (\theta - \rho +\pi/2) +ik_z z }\nonumber\\
   &=& \sum_{k_{\perp},k_z} \sum_{\ell} e^{i\ell \pi/2} A_{\ell}(k_{\perp}, k_z)J_{\ell}(k_{\perp}r) e^{i(\ell\theta + k_z z)},\label{c301}
   \end{eqnarray}
 where $J_{\ell}(k_{\perp}r)$ is the Bessel function of the first kind.
Thus
the physical quantities in a cylindrical system  are expressed as:
\begin{equation}
\left(\begin{array}{c}
n_{\alpha}(r, \theta, z, t)\\
v_{\alpha r}(r, \theta, z, t)\\
v_{\alpha \theta}(r, \theta, z, t)\\
v_{\alpha z}(r, \theta, z, t)\\
\varphi_{\alpha}(r, \theta, z, t)
\end{array}\right) = \left(\begin{array}{c}
n_{0}\\
0\\
0\\
0\\
0
\end{array}\right) + \sum_{k_{\perp}, k_z,\ell} \int \frac{d\omega}{2\pi}\left(\begin{array}{c}
n_{\alpha \ell}(k_{\perp}, k_z, \omega)\\
v_{\alpha r\ell}(k_{\perp}, k_z, \omega)\\
v_{\alpha \theta \ell}(k_{\perp}, k_z, \omega)\\
v_{\alpha z\ell}(k_{\perp}, k_z, \omega)\\
\varphi_{\alpha}(k_{\perp}, k_z, \omega)
\end{array}\right)J_{\ell}(k_{\perp}r)e^{i\ell \pi/2}e^{i(k z+\ell \theta -\omega t)}.\label{c302}
\end{equation}
A set of linearized equations are obtained from Eqs.~(\ref{c206})-(\ref{c208}) as follows:
\begin{eqnarray}
&&i\omega J_{\ell}(k_{\perp}r)v_{\alpha r\ell}+{\hat \Omega}_{\alpha}J_{\ell}(k_{\perp}r) v_{\alpha \theta \ell}
=v_{T\alpha}^2  \frac{\partial J_{\ell}(k_{\perp}r)}{\partial r}\left(\varphi_{\alpha\ell}
 +\frac{n_{\alpha \ell}}{n_{\alpha 0}}\right)
+k_z\Omega_{\alpha}\rho_{\alpha}^2\frac{\ell}{r}J_{\ell}(k_{\perp}r)v_{\alpha z\ell}, \label{c303}\\
&&i\omega J_{\ell}(k_{\perp}r)v_{\alpha \theta \ell}-{\hat \Omega}_{\alpha}J_{\ell}(k_{\perp}r)v_{\alpha r\ell}
=iv_{T\alpha}^2\frac{\ell}{r}J_{\ell}(k_{\perp}r) \left(\varphi_{\alpha\ell}+\frac{n_{\alpha \ell}}{n_{\alpha 0}}\right)
+ik_z\Omega_{\alpha}\rho_{\alpha}^2\frac{\partial J_{\ell}(k_{\perp}r)}{\partial r}v_{\alpha z\ell}, \label{c304}\\
&&\omega J_{\ell}(k_{\perp}r)v_{\alpha z\ell}
=k_zv_{T\alpha}^2 \left(\varphi_{\alpha\ell} + \frac{n_{\alpha \ell}}{n_{\alpha 0}}\right)-k_z\Omega_{\alpha}\rho_{\alpha}^2 \left[\frac{1}{r}\frac{\partial rJ_{\ell}(k_{\perp}r)}{\partial r}(v_{\alpha \theta \ell})-i\frac{\ell}{r}J_{\ell}(k_{\perp}r) v_{\alpha r\ell}\right],\label{c305}~~~~~
\end{eqnarray}
where
\begin{equation}
{\hat \Omega}_{\alpha}=\Omega_{\alpha}\left[1-\left(\frac{1}{2}k_{\perp}^2 + k_z^2\right)\rho_{\alpha}^2 \right],\label{c306}
\end{equation}
and we have used the relation
\begin{equation}
\nabla_{\perp}^2 J_{\ell}(k_{\perp}r) + k_{\perp}^2J_{\ell}(k_{\perp}r)=0.\label{c307}
\end{equation}
Here, the gyro-viscosity is re-normalized into the cyclotron frequency to reduce the magnetic field. The Larmor motion driven by the local potential makes the neighboring parts of the fluid experience the different phase of the potential and is to lose the ordered motion during the time scale $\omega^{-1}$ of the local potential.
This is because the plasma particles lose their memory of transverse ordered velocity in a time $\omega^{-1}$, during which they are displaced by a distance of the order of the Larmor radius.
Equations (\ref{c303})-(\ref{c305}) are rewritten as
\[
J_{\ell}v_{\alpha r\ell}=-i\frac{v_{T\alpha}^2}{\omega^2-{\hat\Omega}_{\alpha}^2}\left[ \left(\omega\frac{\partial}{\partial r}-{\hat \Omega}_{\alpha}\frac{\ell}{r}\right)J_{\ell}
+\frac{\Omega_{\alpha}}{\omega}k_z^2\rho_{\alpha}^2U_{\alpha}(k_{\perp},k_z,\omega)\left(\omega \frac{\ell}{r}-{\hat \Omega}_{\alpha}\frac{\partial}{\partial r}\right) J_{\ell} \right]
\]
\begin{equation}
\times
\left(\varphi_{\alpha\ell}+\frac{n_{\alpha \ell}}{n_{\alpha 0}}\right), \label{c308}
\end{equation}
\[
J_{\ell}v_{\alpha \theta \ell}= \frac{v_{T\alpha}^2}{ \omega^2-{\hat\Omega}_{\alpha}^2}\!\left[\!\left(\!\omega\frac{\ell}{r}-{\hat \Omega}_{\alpha}\frac{\partial}{\partial r}\!\right)J_{\ell}+\frac{\Omega_{\alpha}}{\omega}k_z^2\rho_{\alpha}^2U_{\alpha}(k_{\perp}, k_z, \omega)\left(\!\omega\frac{\partial}{\partial r}-
{\hat \Omega}_{\alpha}\frac{\ell}{r}\!\right)J_{\ell}\right]
\]
\begin{equation}
\left(\!\varphi_{\alpha\ell}+\frac{n_{\alpha \ell}}{n_{\alpha 0}}\!\right)\!\!, \label{c309}
\end{equation}
\be
J_{\ell}v_{\alpha z\ell}=\frac{k v_{T\alpha}^2}{\omega}U_{\alpha}(k_{\perp}, k_z, \omega)\left(\varphi_{\alpha\ell}+\frac{n_{\alpha \ell}}{n_{\alpha 0}}\right),\label{c310}
\end{equation}
where
\begin{equation}
U(k_{\perp}, k_z, \omega)=\frac{\omega^2-{\hat \Omega}_{\alpha}^2-\Omega_{\alpha}{\hat \Omega}_{\alpha}k_{\perp}^2\rho_{\alpha}^2}{\omega^2-{\hat\Omega}_{\alpha}^2-\Omega_{\alpha}^2k_{\perp}^2k_z^2\rho_{\alpha}^4}.\label{c311}
\end{equation}
Substituting Eqs.~(\ref{c308})-(\ref{c310}) into the continuity equation (\ref{c209}) leads to
\begin{equation}
\frac{n_{\alpha\ell}}{n_{\alpha 0}}=-\frac{V_{\alpha}(k_{\perp}, k_z, \omega)}{1+V_{\alpha}(k_{\perp}, k_z, \omega)} \varphi_{\alpha\ell},\quad \mbox{or} \quad
\varphi_{\alpha\ell} + \frac{n_{\alpha\ell}}{n_{\alpha 0}}=\frac{1}{1+V_{\alpha}(k_{\perp}, k_z, \omega)} \varphi_{\alpha\ell},\label{c312}
\end{equation}
where
\begin{equation}
V_{\alpha}(k_{\perp}, k_z, \omega)=-\frac{k_{\perp}^2v_{T\alpha}^2}{\omega^2-{\hat \Omega}_{\alpha}^2}\left[1-\frac{\Omega_{\alpha}{\hat \Omega}_{\alpha}k_z^2\rho_{\alpha}^2}{\omega^2}U_{\alpha}(k_{\perp}, k_z, \omega)\right] -\frac{k_z^2v_{T\alpha}^2}{\omega^2}U_{\alpha}(k_{\perp}, k_z, \omega).\label{c313}
\end{equation}
The electric displacement ${\bm D}$ is obtained from Eq.~(\ref{c205}) together with Eqs.~(\ref{c308})-(\ref{c310}) and (\ref{c321}) by
\begin{equation}
{\bm D}_{\ell} = \left( \begin{array}{ccc}K_{\perp} & -iK_{\perp 2} & 0\\ iK_{\perp 2} & K_{\perp} & 0\\
0 & 0 & k_z\end{array}\right)\left(\begin{array}{c} E_{r\ell}\\ E_{\theta \ell}\\ E_{z\ell}\end{array}\right),\label{c314}
\end{equation}
where
\begin{eqnarray}
K_{\perp}&=&1-\sum_{\alpha}\frac{\omega_{p\alpha}^2}{\omega^2-{\hat \Omega}_{\alpha}^2}\frac{1}{1+V_{\alpha}}\left(1-\frac{\Omega_{\alpha}{\hat \Omega}_{\alpha}}{\omega^2}k_z^2\rho_{\alpha}^2U_{\alpha}\right),\label{c315}\\
K_{\perp 2}&=& \sum_{\alpha}\frac{\omega_{p\alpha}^2}{\omega^2-{\hat \Omega}_{\alpha}^2}\frac{1}{1+V_{\alpha}}\left(\frac{{\hat \Omega}_{\alpha}}{\omega}-\frac{\Omega_{\alpha}}{\omega}k_z^2\rho_{\alpha}^2U_{\alpha}\right),\label{c316}\\
K_z&=&1-\sum_{\alpha}\frac{\omega_{p\alpha}^2}{\omega^2}\frac{U_{\alpha}}{1+V_{\alpha}}.\label{c317}
\end{eqnarray}
Then Eq.~(\ref{c205}) gives
\[
{\bm \nabla}\cdot{\bm D}=i\left(-K_{\perp}{\bm \nabla}_{\perp}^2 + k_z^2K_z\right)J_{\ell}(k_{\perp}r)=i\left(k_{\perp}^2K_{\perp} + k_z^2K_z \right)J_{\ell}(k_{\perp}r)=0,
\]
from which the Bessel function is shown an eigenfunction and a subsidiary condition is given by
\begin{equation}
k_{\perp}^2K_{\perp} + k_z^2K_z=0, \label{c318}
\end{equation}
which determines the relation between $k_{\perp}$ and $k_{z}$.

The plasma considered here is a singly ionized fullerene plasma where masses and the temperatures of positive and negative ions are the same: $\Omega_{\alpha}^2=\Omega^2$, $\omega_{p\alpha}^2=\omega_p^2$,  and $v_{T\alpha}=v_T$. In this case $U_{\alpha}$ and $V_{\alpha}$ do not depend on $\alpha$ and Eq.~(\ref{c318}) is transformed with Eqs.~(\ref{c315}) and (\ref{c317}) into
\[
k_{\perp}^2K_{\perp}+k_z^2K_z=k_{\perp}^2+k_z^2 +\sum_{\alpha}\frac{\omega_{p\alpha}^2}{v_{T\alpha}^2}\frac{V_{\alpha}}{1+V_{\alpha}}=k_{\perp}^2+k_z^2 + \frac{2\omega_p^2}{v_T^2}\frac{V}{1+V}=0,
\]
which gives
\begin{equation}
V=-\frac{k^2\lambda^2}{2+k^2\lambda^2}=-\frac{k_{\perp}^2v_T^2}{\omega^2-{\hat \Omega}^2}\left(1-\frac{\Omega{\hat\Omega}}{\omega^2}k_z^2\rho^2U\right)-\frac{k_z^2v_T^2}{\omega^2}U,\label{c319}
\end{equation}
 where $k^2=k_{\perp}^2+k_z^2$ and $\lambda^2=v_T^2/\omega_p^2$. The subsidiary condition Eq.~(\ref{c318}) may be replaced by Eq.~(\ref{c319}). Here $K_{\perp}, K_{\perp 2}$ and $K_z$ are rewritten as
 \begin{eqnarray}
K_{\perp}&=&1-\frac{2\omega_{p}^2+k^2v_T^2}{\omega^2-{\hat \Omega}^2}\left(1-\frac{\Omega{\hat \Omega}}{\omega^2}k_z^2\rho^2U\right),\label{c320}\\
K_{\perp 2}&=& \frac{2\omega_{p}^2+k^2v_T^2}{\omega^2-{\hat \Omega}^2}\left(\frac{{\hat \Omega}}{\omega}-\frac{\Omega}{\omega}k_z^2\rho^2U\right),\label{c321}\\
K_z&=&1-\frac{2\omega_{p}^2+k^2v_T^2}{\omega^2}U.\label{c322}
\end{eqnarray}

 Since the plasma is detached from the wall, the boundary conditions are that the normal displacement ($D_{r, in}=D_{r, out}$) and tangential electric field ($E_{\theta, in}=E_{\theta, out}$) are continuous at the plasma boundary $r=r_{\ast}$.
Outside the plasma the potential is determined by the Laplace equation ${\bm \nabla}^2\phi_{out}=0$ with the condition $\phi_{out}(R)=0$ at the wall $r=R$, and the solution is given by
\begin{equation}
\phi_{out}(r) = A\left[K_{\ell}(k_zR)I_{\ell}(k_zr)-I_{\ell}(k_zR)K_{\ell}(k_zr)\right],\label{c323}
\end{equation}
where $K_{\ell}$ and $I_{\ell}$ are the modified Bessel functions of the first and second kind.
 The continuity conditions of the normal displacement and the tangential electric field are
 given by
 \begin{equation}
 K_{\perp}\frac{\partial J_{\ell}(k_{\perp}r_{\ast})}{\partial r}+K_{\perp 2}\frac{\ell}{r_{\ast}}J_{\ell}(k_{\perp}r_{\ast})=A\left[K_{\ell}(k_z R)\frac{\partial I_{\ell}( k_z r_{\ast})}{\partial r}-I_{\ell}(k_zR)\frac{\partial K_{\ell}(k_zr_{\ast})}{\partial r}\right], \label{c324}
 \end{equation}
and
\begin{equation}
J_{\ell}(k_{\perp}r_{\ast})=A\left[K_{\ell}(k_zR)I_{\ell}(k_zr_{\ast})-I_{\ell}(k_zR)K_{\ell}(k_zr_{\ast})\right].\label{c325}
\end{equation}
Eliminating $A$ from Eqs.~(\ref{c324}) and (\ref{c325}) the boundary condition is given by
\begin{equation}
K_{\perp}k_{\perp}\frac{J'_{\ell}(k_{\perp}r_{\ast})}{J_{\ell}(k_{\perp}r_{\ast})}-\frac{\ell}{r_{\ast}}K_{\perp 2}=k_z\frac{\phi'_{out}(k_zr_{\ast})}{\phi_{out}(k_zr_{\ast})},\label{c326}
\end{equation}
where
\[ J_{\ell}' = \frac{1}{k_{\perp}}\frac{\partial J_{\ell}(k_{\perp}r)}{\partial r}, \quad  I_{\ell}'=\frac{1}{k_z}\frac{\partial I_{\ell}(k_z r)}{\partial r}=, \quad K_{\ell}'=\frac{1}{k_z}\frac{\partial K_{\ell}(k_zr)}{\partial r}. \]
In the following we consider the case $\ell=0$ since in the experiment the observed waves are azimuthally symmetric.
Thus the boundary condition (\ref{c326}) becomes
\begin{equation}
K_{\perp}=G(k_{\perp}, k_z),\label{c327}
\end{equation}
where
\[
 G(k_{\perp}, k_z)=\frac{k_z}{k_{\perp}} \frac{K_{0}(k_zR)I_{0}'(k_zr_{\ast})-I_{0}(k_zR)K_{0}'(k_z r_{\ast})}{K_{0}(k_z R)I_{0}(k_zr_{\ast})-I_{0}(k_zR)K_{0}(k_z r_{\ast})  }\frac{J_{0}(k_{\perp}r_{\ast})}{J_{0}'(k_{\perp}r_{\ast})}.
 \]
 Noting the following relation with Eq.~(\ref{c311})
 \[ 1 - \frac{\Omega{\hat\Omega}}{\omega^2}k_z^2\rho^2U(k_{\perp},k_z, \omega)=\frac{\omega^2-{\hat\Omega}^2}{\omega^2}\frac{\omega^2-\Omega{\hat \Omega}k_z^2\rho^2-\Omega^2k_z^2k_{\perp}^2\rho^4}{\omega^2-{\hat\Omega}^2-\Omega^2k_z^2k_{\perp}^2\rho^4}, \]
 we obtain  from Eq.~(\ref{c320})
 \[ K_{\perp}=1-(2\omega_p^2+k^2v_T^2)\frac{\omega^2-\Omega{\hat\Omega}k_z^2\rho^2-\Omega^2k_z^2k_{\perp}^2\rho^4}{\omega^2-{\hat\Omega}^2-\Omega^2k_z^2k_{\perp}^2\rho^4}, \]
 which leads with Eq.~(\ref{c327}) to the dispersion equation 
\begin{equation}
\frac{\omega^2-(\Omega{\hat \Omega}+\Omega^2k_{\perp}^2\rho^2)  k_z^2\rho^2}{\omega^2(\omega^2-{\hat\Omega}^2-\Omega^2k_z^2k_{\perp}^2\rho^4)}=\frac{1-G}{2\omega_p^2+k^2v_T^2}, \label{c328}
 \end{equation}
which is supplemented by the subsidiary condition Eq.~(\ref{c319}) which reads
\begin{equation}
\frac{k^2\omega^2-k_z^2({\hat \Omega}+\Omega_{\perp}^2\rho^2)^2}{\omega^2(\omega^2-{\hat \Omega}^2-\Omega^2k_z^2k_{\perp}^2\rho^4)}=\frac{k^2}{2\omega_p^2+k^2v_T^2}. \label{c329}
\end{equation}

The dispersion relations are obtained by solving Eq.~(\ref{c328})
 and substituting the solutions into Eq.~(\ref{c329}) to determine $k_{\perp}$ against $k_z$.
Equations (\ref{c328})
 is simply solved to give the upper hybrid wave, cylindrical Langmuir wave, ion cyclotron wave, and ion sound wave
as
\[
\omega_{\pm}^2=\frac{1}{2}\left({\hat \Omega}^2 + \Omega^2k_{\perp}^2k_z^2\rho^4 +\frac{2\omega_p^2+k^2v_T^2}{1-G}\right)
\]
\be
\times \left\{1 \pm \sqrt{1-\frac{4k_z^2\rho^2\left(\Omega{\hat \Omega}+\Omega^2k_{\perp}^2\rho^2\right)\left(2\omega_p^2+k^2v_T^2\right)/(1-G)}{[{\hat \Omega}^2+\Omega^2k_{\perp}^2k_z^2\rho^4+(2\omega_p^2+k^2v_T^2)/(1-G)]^2}}\,\right\}.\label{c330}
\end{equation}
This dispersion relation is discussed in detail in the next section.

\vspace{0.7cm}

\noindent{\bf IV. \,\,\, Linear Eigenmodes in a Pair-Ion Plasma}

\vspace{0.3cm}

The plasma parameters which we use are the same as those used in the experiment; the density $n_0=1 \times 10^8$ cm${}^{-3}$, the mass $m_i=720 m_p$ ($\omega_p/2\pi$ = 78.2 KHz), the magnetic field $B=0.2$ T ($\Omega/2\pi=4.2$ KHz), the temperature $T=0.3$ eV (in the experiment $T \sim 0.3 - 0.5$  eV), the plasma radius $r_{\ast}=1.5$ cm, and the vessel radius $R=4$ cm. Thus the Debye length $\lambda=$0.04 cm, the Larmor radius $\rho=0.75$ cm, and the thermal speed $v_T=$200 m/s.

The dispersion relations observed in the experiments are given in Fig.~1 where
the blue dots are data obtained for waves produced by a grid exciter, the red ones are those by a cylindrical exciter.
\begin{figure}[h]
\begin{center}
\includegraphics[height=7cm,bb=0 0 260 183,clip=]{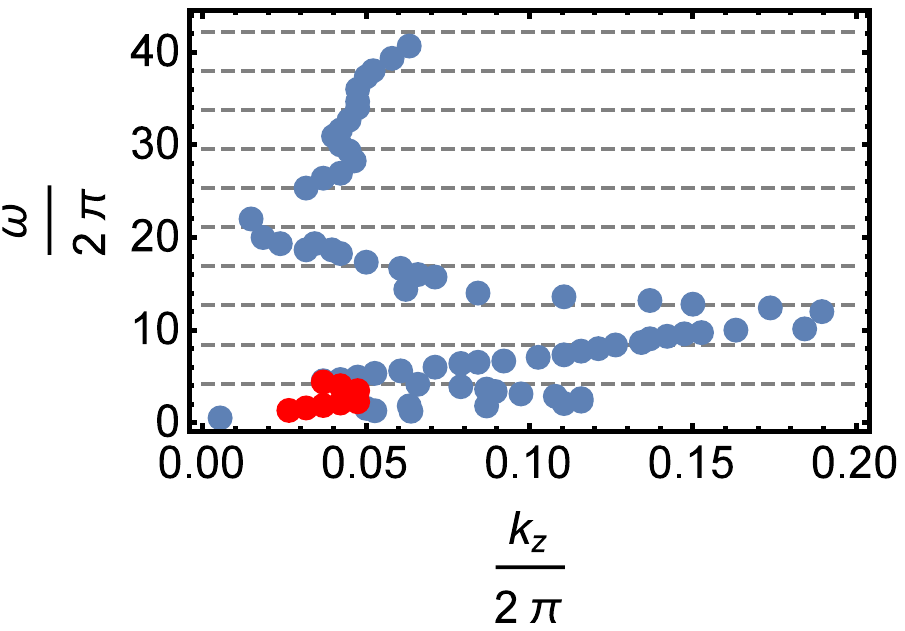}
\end{center}
\caption{Dispersion relations ($\omega/2\pi$ vs $k_z/2\pi$)  observed in the experiment. The blue dots are for the waves excited  by a grid exciter and the red ones are by a cylindrical exciter.}
\end{figure}
Regarding the observed dispersion curves, it has been shown  by the present authors $^{6-7}$ by using the kinetic theory that the low frequency waves are attributed to the ion sound wave (forward wave) and the ion cyclotron wave (backward wave). The high frequency wave is shown to be the cylindrical ion Langmuir wave, while the upper hybrid wave is far beyond the frequency range observed in the experiment. The intermediate frequency waves in the range of $\Omega \le \omega \le 5\Omega$ are results of coupling among the even ($2m$) and odd ($2m+1$) harmonic ion cyclotron waves with the mode number $m \ge 1$.

Thus the kinetic theory is successful since it naturally takes into account the finite Larmor radius effects and higher ion cyclotron resonances, but
the kinetic derivations are so involved. The present fluid theory, however,  provides  a simple and transparent structure to the dispersion relations since the gyro-viscosity is renormalized into
the ion cyclotron frequency which itself indicates the backward nature of dispersion. A disadvantage of a fluid theory which treats only fundamental modes of eigen-waves excited in a system and is not able to describe waves associated with the higher cyclotron resonances is compensated by simple derivations and clear picture based on the renormalization.

\begin{figure}[h]
\begin{center}
\includegraphics[height=7cm,bb=0 0 260 180,clip=]{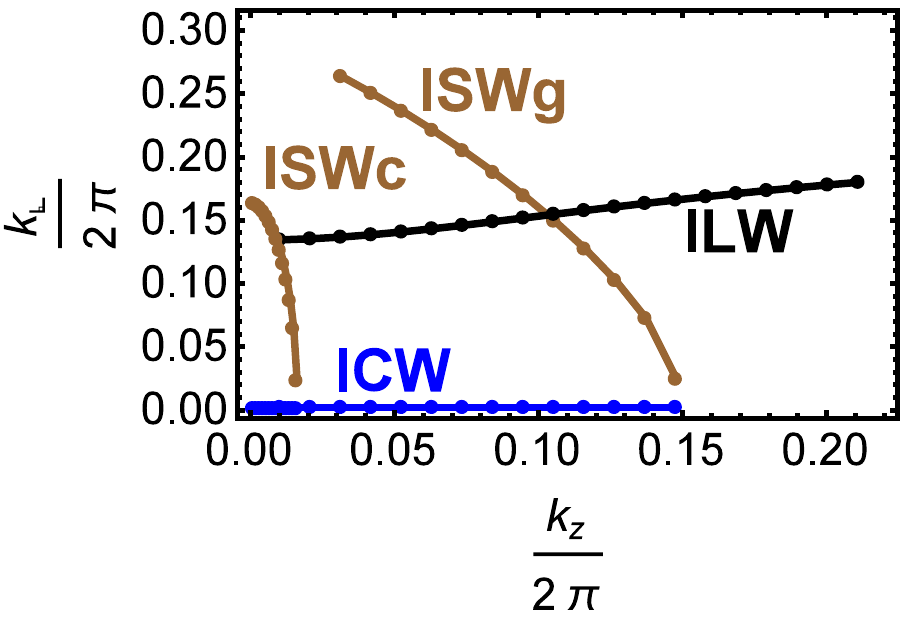}
\end{center}
\caption{Discretized $k_{\perp}/2\pi$ against $k_z/2\pi$ determined by Eq.~(\ref{c329}). The unit of the wavenumber is $cm^{-1}$.}
\end{figure}

The dispersion relation Eq.~(\ref{c330}) has basically four different solutions, i.e., the ion sound, ion cyclotron, upper hybrid,  and cylindrical ion Langmuir waves depending on the set of the wavenumbers $(k_{\perp}, k_z)$ for which $G(k_{\perp}, k_z)$ periodically changes with respect to $k_{\perp}$  from $-\infty$ to $\infty$. Each of the modes has several sub branches. However $k_{\perp}$ is limited to $k_{\perp}r_{\ast} \le j_{01}$ where $j_{01}$ is the first zero of the Bessel function $J_0(k_{\perp}r_{\ast})$ since the radial profile of the observed potential does not have nodes,
\begin{equation}
 \quad \frac{k_{\perp}}{2\pi} \le 0.25516. \label{c401}
\end{equation}
Under the condition Eq.~(\ref{c401}), the sub branches belonging to each mode are mostly prohibited but $G(k_{\perp}, k_z)$ changes from $-\infty$ to $\infty$.
From Eq.~(\ref{c329}) the dispersion relation is approximately given as
\begin{numcases}
    {\omega_{+}^2 \sim}
    \Omega^2\left[(1-\displaystyle{\frac{1}{2}}k_{\perp}^2\rho^2 -k_z^2\rho^2)^2 +k_{\perp}^2k_z^2\rho^4\right], & for $(k_{\perp}, k_z)\in$ ICW, \label{c402}\\[0.2cm]
\displaystyle{\frac{2\omega_p^2+k^2v_T^2}{1-G(k_{\perp}, k_z)}}, & for $(k_{\perp}, k_z)\in$ ILW, \label{c403}\\[0.2cm]
{\hat \Omega}^2 + \Omega^2k_{\perp}^2k_z^2\rho^4 +2\omega_p^2+k^2v_T^2, & for $(k_{\perp}, k_z)\in$ UHW, \label{c404}
 \end{numcases}
and
\begin{equation}
\omega_{-}^2 \sim ({\hat \Omega}\Omega + \Omega^2k_{\perp}^2\rho^2)k_z^2\rho^2=k_z^2v_T^2\left[1+(\frac{1}{2}k_{\perp}^2-k_z^2)\rho^2\right], \quad \mbox{for} \quad (k_{\perp}, k_z) \in \mbox{ISW},\label{c405}\\[0.2cm]
\end{equation}
where ICW, ILW, UHW and ISW are the set of the wavenumbers $(k_{\perp}, k_z)$ determined by Eq.~(\ref{c329}) for the ion cyclotron wave, the ion Langmuir wave, the upperhybrid wave and the ion sound wave, respectively.

The frequency of the upper hybrid wave is much higher than the frequencies of the other waves and is out of the frequency range of the experimental studies. In the following we consider the waves whose frequencies are in the range observed in the experiment.
The discretized wavenumbers $k_{\perp}/2\pi$ against $k_z/2\pi$ are shown in Fig.~2 where the colors of the curves correspond to the ones of the calculated dispersion curves in Figs.~3,~4. Here ISWg is for the ion sound wave by the grid exciter and ISWc is for the one by the cylindrical exciter.

\begin{figure}[h]
\begin{center}
\includegraphics[height=7cm,bb=0 0 262 191,clip=]{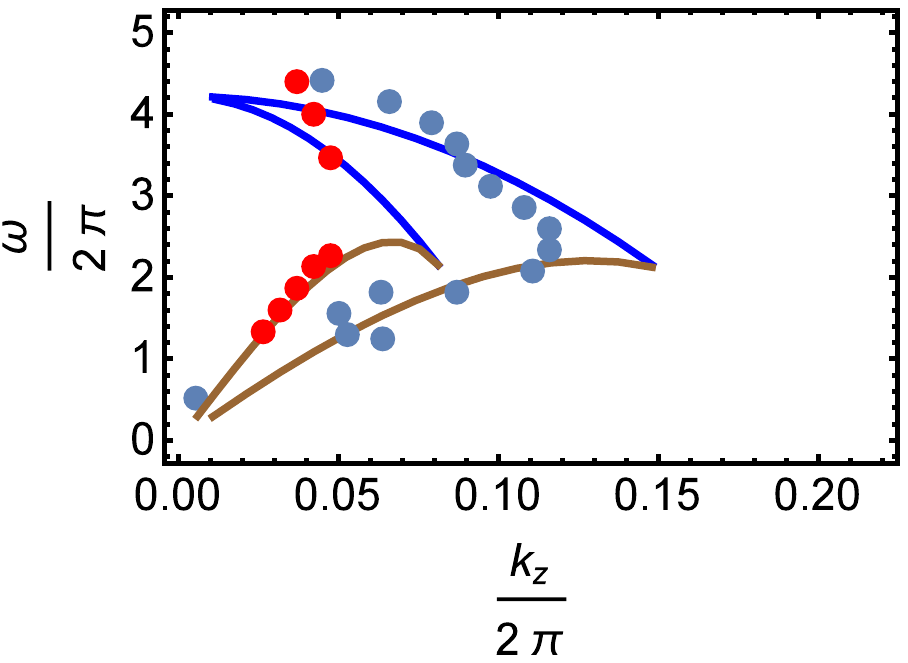}
\end{center}
\caption{Dispersion relations ($\omega/2\pi$ vs $k_z/2\pi$)  of the ion cyclotron wave (the blue curve) and ion sound wave (the brown curve). The plasma parameters mentioned in the text are used to the curve for the blue dots while $T=1$ eV is used to the curve for the red dots with keeping the other parameters the same. The unit of the frequency is KHz.}
\end{figure}

\begin{figure}[h]
\begin{center}
\includegraphics[height=7cm,bb=0 0 260 185,clip=]{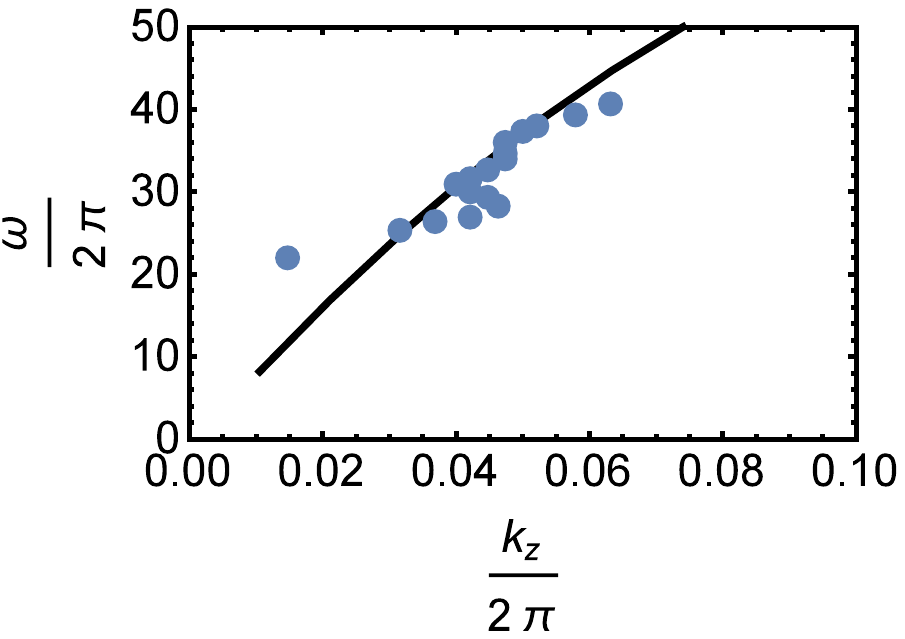}
\end{center}
\caption{Dispersion relation ($\omega/2\pi$ vs $k_z/2\pi$)  of the cylindrical ion Langmuir waves.}
\end{figure}

The ion cyclotron wave and the ion sound wave are coupled through Eq.~(\ref{c329}) and their dispersion relations are terminated at $k_z/2\pi \sim 0.15$ above which $k_{\perp}$ does not exist as is shown in Fig.~3 where the ion cyclotron wave is represented in blue and the ion sound wave is in brown.
The data in the experiment by a grid exciter are well reproduced by Eqs.~(\ref{c402}) and (\ref{c405}) for the parameters mentioned above.  The slope of the dispersion relation of the ion sound wave is the ion thermal speed and Fig.~3 shows clearly that the temperature of the red dots plasma is higher than that of the blue dots plasma. The red dots are  compared with Eqs.~(\ref{c402}) and (\ref{c405})  when the temperature is set to $T =1$ eV and the other parameters remain the same.

The cylindrical ion Langmuir wave Eq.~(\ref{c403}) is shown in Fig.4 where the experimental data may be a mixture of the ion Langmuir wave and higher harmonics of the ion cyclotron wave. The fragments of the higher harmonics of the ion cyclotron wave are beyond the present fluid theory.

The upper hybrid wave given by Eq.~(\ref{c404}) is also realized for  the wave number  $0.25508 \le k_{\perp}/2\pi \le 0.25516$ which satisfies the condition Eq.~(\ref{c401}) but $\omega/2\pi$ is above 110 MHz which is out of the frequency range of the experiment. This is  backward for $k_z/2\pi \le 0.08$ and turns to forward for larger $k_z/2\pi$.  This is a counterpart of the Trivelpiece-Gould mode$^{15}$ in a cylindrically confined electron-ion plasma.

\vspace{0.7cm}

\noindent{\bf VI. \,\,\, SUMMARY AND CONCLUSION }

\vspace{0.7cm}

Although the kinetic theory$^{6-7}$ has been already developed to identify the dispersion relations observed in the experiment, indicating that the finite Larmor radius effects are essential for the excitation of backward waves,  the kinetic theory is so involved in exact particle orbits in phase space, finite Lamor radius effects and higher cyclotron resonances and therefore we have
developed a fluid theory on the propagation characteristics of electrostatic waves in a homogeneous pair-ion plasma. We have taken the gyro-viscosity into account and shown how the ion cyclotron wave is converted  by the gyro-viscosity from a forward wave to a backward wave. Here the  Fourier-Hankel transformation is a vital component in deriving an exact linear dispersion relation and the associated radial eigenfunction in a cylindrical plasma.
For pair-ion plasmas like a fullerene plasma in the experiments,$^{1-5}$
the Larmor radius is large enough for the gyro-viscosity to suppress thermal effects causing the forward nature of the dispersion. It is worthwhile to note that the gyro-viscosity by Braginskii seems to be applicable to plasmas whose scale length is not so large compared with the Larmor radius.

Certainly a fluid theory is unable to describe the higher harmonics of the ion cyclotron wave. However,
the present theory suggests that the intermediate frequency wave as a coupled mode of the higher harmonics of the ion cyclotron waves is expected to be backward as well since the gyro-viscosity is renormalized into ion cyclotron frequency which becomes a decreasing function with respect to $k_z$. In this way the fluid theory provides a clear-cut insight behind the backward nature of the dispersion relations.

ACKNOWLEDGMENTS:
JV acknowledges the financial support from the Spanish
Ministry of Economy and Competitiveness (MINECO)
under the 2011 Severo Ochoa Program MINECO SEV-
2011-0187.\\


\begin{thebibliography}{99}
\bibitem{oh1} W. Oohara and R. Hatakeyama,  Phys. Rev. Lett. {\bf 91}, 205005 (2003).
\bibitem{oh2} W. Oohara, D. Date, and R. Hatakeyama, Phys. Rev. Lett. {\bf 95}, 175003 (2005).
\bibitem{oh3} W. Oohara and R. Hatakeyama, Phys. Plasmas {\bf 14}, 055704 (2007).
\bibitem{oh4} W. Oohara, Y. Kuwabara, and R. Hatakeyama, Phy. Rev. E {\bf 75}, 056403 (2007).
\bibitem{oh5} W. Oohara, T. Hibino, T. Higuchi,  and T. Ohta,  Rev. Sci. Inst. {\bf 83}, 083509 (2012).
\bibitem{kv1} M. Kono, J. Vranjes and N. Batool, Phys. Plasma {\bf 20}, 122111 (2013)
\bibitem{kv2} M. Kono, J. Vranjes and N. Batool, Phys. Rev. Lett. {\bf 112}, 105001 (2014)
\bibitem{v1} J. Vranjes and S. Poedts, Plasma Sources Sci. Technol. {\bf 14}, 485 (2005).
\bibitem{ver1} F. Verheest, Phys. Plasmas {\bf 12}, 032304 (2005).
\bibitem{sal1} H. Saleem, J. Vranjes, and S. Poedts, Phys. Lett. A {\bf 350}, 375 (2006).
\bibitem{kems} I. Kourakis, A. Esfandyari-Kalejahi, M. Medhipoor, and P. K. Shukla. Phys. Plasmas {\bf 13}, 052117 (2006).
\bibitem{zh} B. Zhao and J. Zheng, Phys. Plasmas {\bf 14}, 062106 (2007).
\bibitem{v3} J. Vranjes, D. Petrovic, B. P. Pandey, and S. Poedts, Phys. Plasmas {\bf 15}, 072104 (2008).
\bibitem{v4} J. Vranjes and S. Poedts, Phys. Plasmas {\bf 15}, 044501 (2008).
\bibitem{tri}  A. W. Trivelpiece and R. W. Gould,  J. Appl. Phys. {\bf 30}, 1784 (1959).



\end{thebibliography}
\end{document}